\documentclass[3p,onecolumn]{elsarticle}
\usepackage{amssymb}
\usepackage[active]{srcltx}

\newcommand{\bea}{\begin{eqnarray}}
\newcommand{\eea}{\end{eqnarray}}
\newcommand{\bean}{\begin{eqnarray*}}
\newcommand{\eean}{\end{eqnarray*}}

\def\W #1{\widetilde{#1}}

\def\a{{\alpha}}

\def\eps{\epsilon}

\def\Label{\label}

\begin{document}

\title{Solving for tadpole coefficients in one-loop amplitudes}
\author[cea]{Ruth Britto}
\author[cms,cias]{Bo Feng}
\address[cms]{Center of Mathematical Science, Zhejiang University, Hangzhou, China}
\address[cias]{Division of Applied Mathematical and Theoretical Physics, \\
China Institute for Advanced Study, \\ Central University of Finance
and Economics, Beijing, 100081, China}
\address[cea]{Institut de Physique
Th\'eorique, CEA/Saclay, 91191 Gif-sur-Yvette-Cedex, France}

\begin{abstract}
One-loop amplitudes may be expanded in a basis of scalar integrals
multiplied by  rational coefficients.  We relate the coefficient of
the one-point integral to the coefficients of higher-point
integrals, by considering the effects of introducing an additional,
unphysical propagator, subject to certain conditions.
\end{abstract}
%

\maketitle

\section{Introduction}

One-loop scattering amplitudes may be expanded in a sum of scalar
integrals~\cite{pv,reduction,MasterIntegrals} multiplied by rational
coefficients.  This expansion arises explicitly in typical
computational approaches, reviewed for example in \cite{Bern:2008ef}.  The coefficients may be derived directly
by reduction of Feynman integrals \cite{pv,reduction}, or they may be sought  as
solutions to linear equations taken from various singular limits,
such as unitarity cuts, in an on-shell formalism \cite{unitarity}.  Within the second
approach, the coefficients can be found by
applying ``generalized unitarity'' multi-cuts \cite{multicuts,Ossola:2006us,Ellis:2008ir}.  Alternatively, since the master
integrals are known explicitly and feature unique (poly)logarithms,
they can also be distinguished by the usual unitarity cuts, which
are double-cuts \cite{vanNeerven:1985xr}.  One way to do this is by rewriting the measure of
the cut integral in spinor variables, and then applying the residue
theorem \cite{spinorint}.

This procedure of spinor integration has been carried out in generality for
renormalizable theories with arbitrary massless particles and massive scalars, and analytic expressions for
the coefficients of the scalar pentagon/box, triangle, and bubble
integrals have been given \cite{coefficients}.  However, the tadpole coefficients are missing,
simply because they are obviously free of cuts in physical channels.
Our note addresses this point.

We find that we can solve for the tadpole coefficients in terms of
the  coefficients of higher-point integrals after introducing an
auxiliary, unphysical propagator.  The auxiliary loop integral then
has two propagators, so we can apply unitarity cuts formally.  The
tadpole coefficient is accordingly related to the bubble coefficient
of the auxiliary integral.  Our result is a set of relations giving
the tadpole coefficients in terms of the bubble coefficients of both
the original and auxiliary integrals, and the triangle coefficients
of the auxiliary integrals.  It is interesting to consider whether
this construction might have other applications.

To derive the relations between tadpole coefficients and the others,
we make use of work of Ossola, Papadopoulos, and Pittau (OPP) \cite{Ossola:2006us}, which
gives the result of one-loop reduction at the {\em integrand} level, building upon analysis of their tensor structure \cite{opp0}.
In addition to the integrands for scalar boxes, triangles, bubbles,
and tadpoles, there are a number of ``spurious terms'' which vanish
after integration.  The complete decomposition and classification
given by OPP allows us to relate the original loop integral to the
auxiliary integral including the unphysical propagator.  We then
derive relations among their respective coefficients, and identify
conditions that almost completely decouple the effect of the
unphysical propagator.

We note that on-shell approaches to loop amplitudes face important
subtleties in seeking tadpole coefficients analytically.  The
operation of making a single cut relates  an $n$-point loop
amplitude to an $(n+2)$-point tree level quantity, which should be
considered as an off-shell current.  These are the same starting
points as in proposals to reconstruct full amplitudes entirely from
single cuts \cite{Catani:2008xa,NigelGlover:2008ur}.  Another cut-free
integral is the 0-mass scalar bubble. For applications to physical amplitudes
using unitarity methods, it will be necessary to account for cuts of
self-energy diagrams \cite{Ellis:2008ir}.    In this note, we assume that
these contributions are available. One possibility is to compute cut-free bubble and tadpole contributions analytically by taking careful limits of vanishing mass.
Another
proposal \cite{Bern:1995db} is to fix the tadpole and massless bubble
contributions by universal divergent behavior, once all other
integral coefficients are known.

\section{Relations among cuts and coefficients}

We adopt the notation of OPP \cite{Ossola:2006us,opp0}.  The $D$-dimensional loop momentum is denoted by $\bar q$, whose
4-dimensional component is $q$. The  denominator factors take the
form  $\bar D_i = (\bar q+p_i)^2 - M_i^2$,
where $i=0,1,2,\ldots$.  The tadpole of interest shall be associated
to the factor with $i=0$. We define $\W\ell=-q-p_0.$ and $ K_i
\equiv p_i - p_0. $ Expanding the loop momentum variable into its four-dimensional component plus the remaining
part $\W q$ satisfying $\W q^2 = -\mu^2$, the denominators can be
rewritten as $ \bar D_i = (\W\ell-K_i)^2 - M_i^2 - \mu^2$.

We are interested in the effect of including an {\em auxiliary}
denominator factor,  which we write as
\bea \bar D_K &=&  (\W\ell-K)^2 - M_K^2 - \mu^2.
   ~~~\Label{K-prop} \eea
At this point, $K$ and $M_K^2$
are variables unrelated to the physical amplitude.  Later, they will
 be chosen subject to conditions that minimize the effect of this auxiliary factor.

The one-loop integrand is
\bea I_{true} =  {N(q) \over \bar D_0 \bar D_1 \cdots \bar D_{m-1}}
~~~\Label{true-integrand} \eea
where, following \cite{Ossola:2006us}, we use $N(q)$ to denote the numerator, which
is  a polynomial in $q$. We call the integrand $I_{true}$ the
``true'' integrand to distinguish it from the ``auxiliary''
integrand, which we construct by inserting the auxiliary factor
$\bar D_K$, as follows.
\bea I_K
 = {N(q) \over
\bar D_K \bar D_0 \bar D_1 \cdots \bar D_{m-1}}
~~~\Label{aux-integrand} \eea

Consider the single-propagator cut  of the tadpole of interest.  It
is  the result of eliminating the denominator factor $\bar D_0$ from
the integrand:
\bea
 I^{\rm tree}_{1-cut} = {N(q) \over \bar D_1 \cdots \bar D_{m-1}}
~~~\Label{1-cut-amp} \eea
This integrand is the analog of the product of tree amplitudes   $
A^{\rm tree}_{\rm Left} A^{\rm tree}_{\rm Right}$ obtained from a
standard unitarity cut.  However, from a {\em single} cut, we obtain
a tree amplitude at a singular point in phase space, since two
external on-shell momenta are equal and opposite.  This singularity
can create difficulties that we do not address generally here.  It
is probably best considered as an off-shell current.

In the OPP method \cite{Ossola:2006us}, the integrand is expanded in terms of the master
integrals multiplied by their coefficients in the amplitude, plus
additional ``spurious terms'' which vanish upon integration.
 The
unintegrated expansion is
\bea I_{true} &=& \sum_{i}^{m-1} [a(i)+\W
a(q;i)]I^{(i)} + \sum_{i<j}^{m-1} [b(i,j)+\W b(q;i,j)]I^{(i,j)} +
\sum_{i<j<r}^{m-1} [c(i,j,r)+\W c(q;i,j,r)]I^{(i,j,r)} \nonumber \\
& & + \sum_{i<j<r<s}^{m-1} [d(i,j,r,s)+\W
d(q;i,j,r,s)]I^{(i,j,r,s)}+\sum_{i<j<r<s<t}^{m-1}
 e(i,j,r,s,t) I^{(i,j,r,s,t)} ~~\Label{true-opp-expansion} \eea
where $a(i), b(i,j), c(i,j,r),
d(i,j,r,s), e(i,j,r,s,t)$ are the coefficients of the master
integrals; $\W a(q;i), \W b(q;i,j)$, $\W c(q;i,j,r)$, $\W
d(q;i,j,r,s)$ are the spurious terms which integrate to zero;  and
the master integrals are
\bean I^{(i)} = {1 \over \bar D_i}, \quad I^{(i,j)}= {1 \over
\bar D_i \bar D_j}, \quad I^{(i,j,r)}= {1 \over \bar D_i \bar D_j
\bar D_r},
\quad  I^{(i,j,r,s)} =  {1 \over \bar D_i \bar D_j \bar D_r \bar
D_s},
\quad
I^{(i,j,r,s,t)}= {1 \over \bar D_i \bar D_j \bar D_r \bar D_s \bar D_t}.
\eean
Notice that we have included the pentagon explicitly.  We shall
perform our  analysis of the coefficients in $D=4-2\eps$ dimensions.

Now, consider the auxiliary integrand $I_K$.  On one hand, it is
simply the true integrand divided by the auxiliary propagator. Thus,
using (\ref{true-opp-expansion}), we get the following expansion in master integrals:
 \bea I_K = {I_{true} \over \bar D_K}
 &=& \sum_{i}^{m-1} [a(i)+\W a(q;i)]I^{(K,i)}
+ \sum_{i<j}^{m-1} [b(i,j)+\W b(q;i,j)]I^{(K,i,j)} +
\sum_{i<j<r}^{m-1} [c(i,j,r)+\W c(q;i,j,r)]I^{(K,i,j,r)} \nonumber
\\ & & + \sum_{i<j<r<s}^{m-1} [d(i,j,r,s)+\W
d(q;i,j,r,s)]I^{(K,i,j,r,s)} +\sum_{i<j<r<s<t}^{m-1}
 e(i,j,r,s,t) I^{(K,i,j,r,s,t)}~~\Label{divided} \eea
Notice here that the ``spurious'' terms such as $\int \W b(q;i,j)$
are no  longer spurious with the factor $\bar D_K$ included.  For
example,  while $\int \W b(q;i,j)I^{(i,j)}=0$ by construction, in
general $\int \W b(q;i,j)I^{(K,i,j)}\neq 0$.
 On the other hand, the auxiliary integrand $I_K$ has its own OPP
expansion, where we label the auxiliary coefficients and spurious
terms  by the subscript $K$, and we separate the auxiliary
propagator explicitly, so it is not included in the summation
indices:
\bea I_K
 &=&\sum_{i}^{m-1} [a_K(i)+\W a_K(q;i)]I^{(i)}
+ \sum_{i<j}^{m-1} [b_K(i,j)+\W b_K(q;i,j)]I^{(i,j)} +
\sum_{i<j<r}^{m-1} [c_K(i,j,r)+\W c_K(q;i,j,r)]I^{(i,j,r)} \nonumber
\\ & & + \sum_{i<j<r<s}^{m-1} [d_K(i,j,r,s)+\W
d_K(q;i,j,r,s)]I^{(i,j,r,s)}+ \sum_{i<j<r<s,t}^{m-1}
 e_K(i,j,r,s,t) I^{(i,j,r,s,t)} \nonumber \\ & & + [a_K(K)+\W
a_K(q;K)]I^{(K)} + \sum_{j}^{m-1} [b_K(K,j)+\W b_K(q;K,j)]I^{(K,j)}
+ \sum_{j<r}^{m-1} [c_K(K,j,r)+\W c_K(q;K,j,r)]I^{(K,j,r)} \nonumber
\\ & & + \sum_{j<r<s}^{m-1} [d_K(K,j,r,s)+\W
d_K(q;K,j,r,s)]I^{(K,j,r,s)}+ \sum_{j<r<s,t}^{m-1}
 e_K(K,j,r,s,t) I^{(K,j,r,s,t)} ~~\Label{aux-opp} \eea
With the subscript $K$, terms such as $\W b_K(q;i,j)$ and $\W
b_K(q;K,j)$  are truly spurious terms in  (\ref{divided}), e.g.
$\int\W b_K(q;K,j,r)I^{(K,j)}=0$.

The purpose of introducing the auxiliary integrand $I_K$ is to give
information about the tadpole coefficient by cutting {\em two}
propagators.  So, we will always choose to cut the auxiliary propagator
$\bar D_K$ along with $\bar D_0$. Restricted to the terms involved in this
cut, the auxiliary integrand (\ref{aux-opp}) is
\bea \left. I_K
\right|_{C_{0K}}
 &=&
  [b_K(K,0)+\W b_K(q;K,0)]I^{(K,0)}
+ \sum_{i}^{m-1} [c_K(K,0,i)+\W c_K(q;K,0,i)]I^{(K,0,i)} \nonumber
\\ & & + \sum_{i<j}^{m-1} [d_K(K,0,i,j)+\W
d_K(q;K,0,i,j)]I^{(K,0,i,j)}+ \sum_{i<j<s}^{m-1}
 e_K(K,0,i,j,s) I^{(K,0,i,j,s)} ~~\Label{aux-opp-restricted} \eea
Similarly, we can restrict our attention to the corresponding subset of terms in (\ref{divided}):
\bea
\left. I_K \right|_{C_{0K}}
 &=&  [a(0)+\W a(q;0)]I^{(K,0)}
+ \sum_{i}^{m-1} [b(0,i)+\W b(q;0,i)]I^{(K,0,i)} + \sum_{i<j}^{m-1}
[c(0,i,j)+\W c(q;0,i,j)]I^{(K,0,i,j)} \nonumber \\ & & +
\sum_{i<j<r}^{m-1} [d(0,i,j,r)+\W d(q;0,i,j,r)]I^{(K,0,i,j,r)}+
\sum_{i<j<r<s}^{m-1}  e(0,i,j,r,s) I^{(K,0,i,j,r,s)}
~~\Label{divided-restricted} \eea

Our plan is to find the tadpole coefficient, $a(0)$, by imposing the
equivalence of (\ref{aux-opp-restricted}) and
(\ref{divided-restricted}) {\em after} completing the cut integral.
After integration, the spurious terms of (\ref{aux-opp-restricted})
simply drop out, as they are designed to do so:
\bea \int_{C_{0K}} I_K & = &
  b_K(K,0) \int_{C_{0K}} I^{(K,0)}
+ \sum_{i}^{m-1} c_K(K,0,i) \int_{C_{0K}} I^{(K,0,i)}\nonumber \\
& &  + \sum_{i<j}^{m-1} d_K(K,0,i,j) \int_{C_{0K}} I^{(K,0,i,j)}+
\sum_{i<j<s}^{m-1} e_K(K,0,i,j,s) \int_{C_{0K}} I^{(K,0,i,j,s)}
~~\Label{int-aux-opp-restricted} \eea
Here, the cut integral is
denoted by $\int_{C_{0K}}$, which indicates that we use the
Lorentz-invariant phase space measure including the factor
$\delta(\bar D_0)\delta(\bar D_K)$.

However, the integration of the formula (\ref{divided-restricted})
is not so straightforward, because the original spurious terms no
longer correspond to the structures of the denominators they
multiply.  So,  we shall view the expression
(\ref{divided-restricted}) as a function of the loop momentum $q$,
and find the coefficients of master integrals
analytically, for each of the spurious terms classified by  OPP.

Keeping in mind that our target is the single number $a(0)$, which
appears as part of the auxiliary bubble coefficient in
(\ref{divided-restricted}), we begin by extracting only the
auxiliary bubble contributions of the various spurious terms,
divided by  their denominators as well as $\bar D_K$.  (The other
non-spurious terms, with $b(0,i)$, $c(0,i,j)$, and $d(0,i,j,r)$,
clearly belong entirely to coefficients of other master integrals,
of which the 4-dimensional pentagon is a linear combination of five
boxes.)

Our result is that there are conditions under which most of the
spurious terms have no effect.  Specifically, for all the propagator
momenta $K_i$ inside $\bar D_i$,  we would like to  take
\bea K \cdot K_i = 0, \quad \forall i;~~~~~M_K^2 = M_0^2 + K^2.
~~~\Label{KM-cond} \eea
%
%
We are free to take (\ref{KM-cond}) as a definition of $M_K^2$,
while the condition for $K$ is clearly nontrivial to satisfy
physically. For the
purposes of defining our construction, we perform a formal
reduction.  Any integrand having five or more propagators has at
least four independent momenta $K_i$ that can be used to expand any
external momentum vector appearing in the numerator to do the reduction.
 For integrands with at most four propagators, there are no more than three
momenta $K_i$, and the condition (\ref{KM-cond}) can be satisfied,
for example by the construction $K_\mu=\eps_{\mu\nu\rho\sigma}
K_1^\nu K_2^\rho K_3^\sigma$.  In practice, we consider our procedure to be formal and analytic and propose to set the products $K \cdot K_i$ identically to zero wherever they appear.

If conditions (\ref{KM-cond}) are satisfied, then we find that only
one of all the spurious terms contributes to the auxiliary bubble
coefficient.  Specifically,
\bea
b_K(K,0) = a(0) + {1
\over 12} \sum_i (K_i^2-M_i^2+M_0^2) \W b_{00}[K_i],
\eea
where  $\W b_{00}[K_i]$ is the coefficient of one of the spurious terms defined in \cite{Ossola:2006us} (and hence it depends on all the details of the original integrand).

A convenient way to constrain  $\W b_{00}[K_i]$ is to identify the
effect of the spurious term on the auxiliary triangle coefficient.
Still imposing the conditions (\ref{KM-cond}),  we {\em repeat} our
analysis of all the OPP spurious terms in
(\ref{divided-restricted}), this time isolating the contributions to
{\em triangle} coefficients.  Fortunately, we find  that only this
same single spurious term has a nonvanishing effect, if we focus on
the terms with $\mu^2$-dependence. (We assume that explicit
$\mu^2$-dependence in the numerator $N(q)$ has been set aside.)  The
result is
\bea
\left. c_K(K,0,i) \right|_{\mu^2}=
\left. b(0,i) \right|_{\mu^2}
+ {K_i^2 \over 3}
\W b_{00}[K_i] , \eea
where $|_{\mu^2}$ means the coefficient of $\mu^2$.

Now we propose the following procedure for finding tadpole coefficients.

\begin{enumerate}

\item Find the single-cut expression  $A^{\rm tree}_{1-cut}$ obtained
by cutting the propagator $\bar D_0$.  Expand the numerator in $\mu^2$,
and work term by term, setting aside these explicit factors of $\mu^2$.

\item Construct the true integrand $I=A^{\rm tree}_{1-cut}/\bar D_0$ and
the auxiliary integrand $I_K=A^{\rm tree}_{1-cut}/(\bar D_K 
\bar D_0)$. It may be convenient at this stage already to  choose $K$ and
$M_K$ to satisfy the conditions (\ref{KM-cond}). Alternatively, they
can be taken as arbitrary variables until the final step.

\item Use the cut integral $\int_{C_{0K}}I_K$ to evaluate the auxiliary bubble
coefficient $b_K(K,0)$ and all the auxiliary triangle coefficients $c_K(K,0,i)$.

\item Use the cut integrals  $\int_{C_{0K_i}}I$ to evaluate all the
true bubble coefficients $b(0,i)$.

\item The tadpole coefficient is given by imposing the conditions (\ref{KM-cond})
 in the following expression.
\bea
 a(0) = b_K(K,0) + \sum_i   {K_i^2-M_i^2+M_0^2  \over 4 K_i^2 }  \left.
 \left[ c_K(K,0,i)- b(0,i) \right] \right|_{\mu^2}.
\eea
This formula is valid term by term, having set aside the original explicit
factors of $\mu^2$ in the numerator $N(q)$.

\end{enumerate}

\section{Contributions to $b_K(K,0)$ from spurious terms}

In this section we will discuss the contributions to $b_K(K,0)$ of
the auxiliary integrand $I_K$ from the expression
(\ref{divided-restricted}),  where the terms have been separated
into the scalar integral coefficients, plus spurious terms as
classified by OPP \cite{Ossola:2006us}. As we have discussed, these terms are no longer
``spurious'' in the same sense, once $\bar D_K$ is included.  (N.B.:
OPP write the expansion with all denominators multiplied through, so
that the ``spurious terms'' for them are the polynomial numerators.
Here, we use ``auxiliary spurious terms'' to refer to the
correesponding terms with all denominators present, including
$\bar D_K$.)

The first contribution is obviously  $a(0)$, which is the tadpole
coefficient  that interests us.  Now we discuss the possible
contributions from the terms with $\W a, \W b,\W c,\W d$ in
(\ref{divided-restricted}). We will see why we choose the decoupling
conditions (\ref{KM-cond}). They arise naturally by considering the
terms of lowest degree.  We have proceeded step by step through all
the spurious terms of \cite{Ossola:2006us}. Our results have been derived in the formalism
of \cite{spinorint} and verified using Passarino-Veltman reduction \cite{pv} as implemented in FeynCalc \cite{Mertig:1990an}.

\begin{itemize}

\item {\bf One-point spurious terms:}
In the simplest case, all spurious terms of this type are linear in
the numerator $ I_{11}= {2\W\ell\cdot R_1 / \bar D_0}$. 
The auxiliary integrand, including
$\bar D_K$ in the denominator, is then $ I_{11}^{\bar D_K}= {2\W\ell\cdot
R_1 / (\bar D_K \bar D_0) }$. It is easy to find the scalar bubble
coefficient from a standard unitarity cut (or alternatively, by
straightforward reduction).  The result is
\bea C[\bar D_0,\bar D_K]=
{ (K\cdot
R_1)(K^2+M_0^2-M_K^2) \over K^2}
.~~~~\Label{I11-D0DK}\eea
 There are four independent ``1-point like'' spurious terms
as given by OPP, i.e., four independent values of $R_1$.
We see that we can decouple all their contributions by imposing the condition
\bea  {K^2+M_0^2-M_K^2}=0~~~~\Label{alpha=0} \eea

\item {\bf Two-point spurious terms:} Spurious 2-point terms can be either
linear or quadratic in loop momentum. In the case of linear
dependence, the auxiliary integrand with $\bar D_K$ has the scalar
bubble coefficient 
\bea C[\bar D_0,\bar D_K] & = & {-(K\cdot K_i)(K\cdot R_1)+K^2(K_i\cdot
R_1)\over (K\cdot K_i)^2- K^2K_i^2}.~~~~\Label{I21-D0DK}\eea
In the spurious terms, $R_1$ takes three possible values of vectors,
called $\ell_7,\ell_8, n$.  These vectors are defined in \cite{Ossola:2006us}; here we only need to use some of their properties. (We use  $K$ as the
auxiliary momentum in the OPP construction of these vectors.) In each of these cases, we
have $R_1 \cdot K_i = 0$.  Moreover, $K \cdot \ell_{7/8}=0$, but $K\cdot n\neq 0$.
 To make this last spurious contribution vanish, we enforce a new decoupling condition:
\bea K\cdot K_i=0.~~~~~\Label{KK1=0} \eea

Now we move on to the quadratic spurious 2-point terms.  There are five such terms.
 For four of them, the auxiliary bubble coefficient vanishes under the two
 decoupling conditions.  The fifth spurious term is $K(q;0,i)$, which can be
 written $(\W\ell\cdot n)^2-{((\W\ell\cdot K_i)^2-
K_i^2 \W\ell^2)/ 3}$.  Its coefficient in the OPP expansion is denoted $\W b_{00}(0,i)$.
After imposing the decoupling conditions, the auxiliary bubble coefficient from this term is
\bea C_{\W b_{00}(0,i)}= { K_i^2+M_0^2-M_i^2 \over 12}.
\eea

Because this spurious term  gives a nonzero contribution under
the decoupling conditions, we must calculate it and subtract its
contribution when we calculate the tadpole coefficient.  For this reason,
we will turn to the auxiliary triangles $c_K(K,0,i)$ in the following section.

\item {\bf Three-point and four-point spurious terms:} All of the auxiliary three-point spurious
 terms decouple after
imposing (\ref{KM-cond}).  There is just one auxiliary four-point
spurious term, and it gives no bubble contribution at all, because
its numerator is linear in the loop momentum.

\end{itemize}

To summarize, have seen that if we impose the conditions
(\ref{KM-cond}), then all contributions from spurious terms will
decouple, except one, whose coefficient is $\W b_{00}$. We have
\bea b_{K}(K,0)= a(0)+\sum_{i} \W b_{00}(0,i) {K_i^2+M_0^2-
M_i^2\over 12}~~~~\Label{a0-eq-1}\eea
where $\W b_{00}(0,i)$ is the coefficient of the spurious term
$K(q;0,i)$ as defined by OPP.

In this analysis, we are assuming a renormalizable theory.  We have assumed that  the power of $\W\ell$ in the numerator is
equal to or less than the number of propagators in the denominator.
In those terms where
the power of $\W\ell$ is {\em strictly} less than the number of propagators,
then we have $\W b_{00}(0,i)=0,~\forall i$. Thus we have
$b_{K}(K,0)= a(0)$, i.e., we get the tadpole coefficient $a(0)$
immediately by calculating the bubble coefficient under the
decoupling conditions.
From terms where the power of $\W\ell$ is equal to the number of propagators,
$\W b_{00}(0,i)\neq 0$, and we need  to compute it.
We have
found that we can use a similar decoupling approach
 to calculate the triangle coefficient $C[\bar D_0,\bar D_K, \bar D_i]$
and extract the corresponding $\W b_{00}(0,i)$. This procedure will be discussed
in the next section.

\section{The calculation of $\W b_{00}(0,i)$}

Recall the
expansion (\ref{divided-restricted}), where we augmented the OPP expansion with the extra factor $\bar D_K$, so that the spurious terms no longer integrate to zero.
 We see that the term $\W b(q;
0,i) I^{(K,0,i)}$ contributes not only to the coefficient of the bubble
$I^{(K,0)}$, but also to the coefficient of the triangle $I^{(K,0,i)}$. Thus
it is possible to find $\W b_{00}(0,j)$ from the evaluation of
the coefficient of triangle $I^{(K,0,i)}$ within $I_K|_{C_{0K}}$.

Just as in the previous section, where we studied all contributions to  coefficient of $I^{(K,0)}$ from
$I_K|_{C_{0K}}$, so $c_K(K,0,i)$ in (\ref{aux-opp-restricted}) also receives
 contributions from the original spurious terms in (\ref{divided-restricted}).
Thus we carry out the corresponding analysis in this section.
We impose the decoupling conditions from the start.
Then we find that three of the auxiliary spurious terms still give auxiliary triangle contributions.
The first nonvanishing contribution comes, indeed, from the term we want, namely  $K(q;0,i)$.  Its auxiliary triangle contribution, after having applied the decoupling conditions, is
\bea C[\bar D_0,\bar D_K,\bar D_i]_{\W b_{00}}= -{(K_1^2+M_0^2-M_i^2)^2-4
K_i^2(M_0^2+\mu^2)\over 12}.
~~~\Label{tri-22}
\eea

The second and third nonvanishing contributions come from the spurious 3-point terms with quadratic dependence on the loop momentum.
The auxiliary integrand is
\bea I_{32}^{\bar D_K} ={(2\W\ell\cdot R_1)^2\over \bar D_0 \bar D_K \bar D_i
\bar D_j},
\eea
where $R_1$ takes two values, called $\ell_{3,4}$.  After applying the decoupling conditions,
  we find that the triangle coefficient is
\bean - {(K\cdot R_1)^2 ((K_i\cdot K_j)
(K_i^2+M_0^2-M_i^2) -K_i^2(K_j^2+M_0^2-M_j^2))\over K^2 ((K_i\cdot
K_j)^2- K_i^2 K_j^2)} \eean
This quantity does not vanish identically, so our decoupling might seem to be inadequate.  But there is good news here:
{\sl the contribution does not depend on $\mu^2$}, while the contribution
from $\W b_{00}$ {\sl does} depend on $\mu^2$.  Thus we can use the $\mu^2$-dependence to find exactly the term we need.

Our plan is now clear: (1) calculate the bubble coefficient $b(0,i)$ from the integrand $I_{true}$, which does not contain
$\bar D_K$; (2) calculate the triangle coefficient $c_K(K,0,i)\equiv C[\bar D_0,
\bar D_K, \bar D_i]$ of the integrand $I_K$, which does contain $\bar D_K$; (3) find the $\mu^2$-dependent
terms in $b(0,i)$ and $c_K(K,0,i)$,  and then solve the following
equation to find $\W b_{00}(0,i)$.
\bea c_K(K,0,i)|_{\mu^2} & = & b(0,i)|_{\mu^2}+ {  K_i^2 \over
3} \W b_{00}(0,i),~~~~\Label{c-K-i} \eea
where $|_{\mu^2}$ means the coefficient of $\mu^2$.
After computing $ \W b_{00}(0,i)$ for every $i$, we substitute back into (\ref{a0-eq-1})
and  finally find $a(0)$, the tadpole coefficient.

\section{Discussion}

In closing, we list some formulas we obtained from our algorithm and comment on their properties. We denote a general integrand term by two indices $n,m$, writing $I_{n,m}[ \{
K_i\}, \{ R_j\}]= {\prod_{j=1}^m (2\W\ell\cdot R_j)/(
\prod_{i=0}^{n-1} \bar D_i)}$, where for additional simplicity we are setting $R_3=R_1$.  Further, we define $\a_i= K_i^2+M_0^2-M_i^2$ and 
$\Delta_{ij}=(K_i\cdot K_j)^2- K_i^2 K_j^2$.  We list results from the first few of these integrands here. 
{\small
\bea a(0)_{I_{21}} & = &  {- R_1\cdot K_1\over K_1^2}~~~\Label{I21-tad}\\
a(0)_{I_{22}} & = & -{ \a_1 [(R_1\cdot R_2) K_1^2- 4(R_1\cdot
K_1)(R_2\cdot K_1)]\over 3 (K_1^2)^2}~~~\Label{I22-tad}
\eea
\bea
a(0)_{I_{32}} & = & { \sum_{i,j=1}^2 A_{ij} (K_i\cdot
R_1)(K_j\cdot R_2)\over K_1^2 K_2^2
\Delta_{12}},
\eea
where
\bea
A_{11}=K_2^2(K_1\cdot K_2), \quad A_{22}=K_1^2(K_1\cdot
K_2), \quad A_{12}=A_{21}=-K_1^2 K_2^2 ~~~\Label{I32-tad}\eea
\bea a(0)_{I_{33}} & = & 
{\sum_{i=1,2} A_{i,00} \left(2(R_1\cdot R_2) R_1\cdot K_i+R_1^2 R_2 \cdot K_i \right)
\over 3 K_i^2  \Delta_{12}}
+
{\sum_{i=1,2} \sum_{j\leq k} A_{i; jk}
(K_i\cdot R_2) (K_j\cdot R_1)(K_k\cdot R_1)\over 3 (K_1^2 K_2^2)^2
\Delta_{12}^2},~~~\Label{I33-tad} \eea
} where
{\small \bean
A_{1;00} &=& 4\alpha_2 K_1^2 -4\alpha_1 K_1 \cdot K_2, \qquad
A_{2;00} = 4\alpha_1 K_2^2 - 4\alpha_2 K_1 \cdot K_2 \\
A_{1;11} & = & -2 (K_2^2)^2(  \Delta_{12}(4 \a_1 (K_1\cdot
K_2)-2\a_2 K_1^2)+5 K_1^2(K_1\cdot K_2)(\a_2 (K_1\cdot K_2)-\a_1
K_2^2))\\
A_{2;11} & = &  A_{1;12} =  -2 K_1^2 (K_2^2)^2(\a_1 \Delta_{12}-5
K_1^2(\a_2 (K_1\cdot K_2)-\a_1 K_2^2)),\\
A_{2;22} & = & A_{1;11}|_{\a_1, K_1\leftrightarrow
\a_2,K_2},~~~~A_{2;12}=A_{1;22}=A_{1;12}|_{\a_1, K_1\leftrightarrow
\a_2,K_2}\eean
\bea a(0)_{I_{43}} & = & { \sum_{i=1}^3\sum_{t\leq s,1}^3 A_{i;ts}
(K_i\cdot R_2) (K_t\cdot R_1)(K_s\cdot R_1)\over K_1^2 K_2^2 K_3^2
\Delta_{12}\Delta_{13}\Delta_{23}\Delta_{123}}, \eea
} where {\small
\bean \Delta_{123} & = & K_1^2 K_2^2 K_3^2+2 (K_1\cdot K_2)(K_2\cdot
K_3)(K_3\cdot K_1)-K_1^2(K_2\cdot K_3)^2-K_2^2(K_1\cdot
K_3)^2-K_3^2(K_1\cdot K_2)^2\\
 A_{1;11}  & = &2 \Delta_{23} K_2^2
K_3^2\left\{ (K_2\cdot K_3)[ (K_1\cdot K_2)^2\Delta_{13}+ (K_1\cdot
K_3)^2\Delta_{12}] +(K_1\cdot K_2)(K_1\cdot K_3)[
K_2^2\Delta_{13}+K_3^2\Delta_{12}]\right\}\\
A_{1;12} & = & A_{2;11}=2 K_1^2 K_2^2
K_3^2\Delta_{13}\Delta_{23}(-K_2^2(K_1\cdot K_3)+(K_2\cdot
K_1)(K_2\cdot K_3)),~~~~~A_{1;13} =  A_{3,11}=
[A_{1;12}]|_{K_2\leftrightarrow K_3},\\
A_{2;22} &= & [A_{1;11}]|_{K_1\leftrightarrow K_2},~~~A_{3;33} =
[A_{1;11}]|_{K_1\leftrightarrow K_3}, ~~~A_{3;23} = A_{2;33}=
[A_{1;12}]|_{K_1\leftrightarrow K_3},~~
A_{2;12}=A_{1;22}= [A_{1;12}]|_{K_1\leftrightarrow K_2},\\
A_{2;23}& = &  A_{3;22}= [A_{1;12}]|_{K_1, K_2, K_3\to K_2,
K_3,K_1},~~~ A_{3;13} =  A_{1;33}=[A_{1;12}]_{K_1,K_2, K_3\to K_3,
K_1,
K_2},\\
A_{1;23} & = & A_{2;13}=A_{3;12}= -2 K_1^2 K_2^2 K_3^2
\Delta_{12}\Delta_{13}\Delta_{23}
\eean
}
We note some  patterns in these
tadpole coefficients: (1) the tadpole coefficient is independent of 
$\mu^2$; (2) for $I_{n,n-1}$, the coefficient is independent of masses; (3) for $I_{n,n}$, the coefficient is of the form
$\sum_i \a_i c_i$.

Finally, we offer a comment on massless limits, for cases involving 0-mass scalar bubble integrals.  These integrals are cut-free, and are in fact linear combinations of tadpoles.  Therefore, it seems we will face another obstacle in determining their coefficients.  We find that nevertheless, we can apply our analytic formalism to bubble coefficients by keeping all appearances of $K_i^2$ throughout the calculation, taking limits of $K_i^2 \to 0$ only at the very end, with tadpoles and massless scalar bubbles combined appropriately.  For example, consider the integrand $I_{21}$.  
 The full reduction of $I_{21}$ is
\bea I_{21} &= & (R_1\cdot K_1) \left( 1+{M_0^2-M_1^2\over
K_1^2}\right) I_{2,0}- {R_1\cdot K_1\over K_1^2}I_{1,0}[\bar D_0]+
{R_1\cdot K_1\over K_1^2} I_{1,0}[\bar D_1].
~~~~\Label{Exa-1-final}\eea
The coefficients of the tadpoles and the scalar bubble diverge individually in the limit $K_1^2 \to 0$.  However, the complete sum is finite in the limit.  The same pattern holds for more complicated integrands. In taking the limit, it is important to use the complete expansion of the scalar bubble integral in the parameter $K_1^2$.  This analysis will be presented in detail elsewhere \cite{bfnext}.  The procedure given in this paper is sufficient to determine tadpole coefficients in terms of bubble and triangle coefficients while the parameters $K_i^2$ are still formally finite.

\section*{Acknowledgments}

We thank the participants of the {\em International workshop on gauge and string amplitudes} at IPPP Durham, where this work has been presented.
The work of R.B. was supported by Stichting FOM;
by Fermilab, which is operated by Fermi Research Alliance, LLC  under Contract No.
DE-AC02-07CH11359 with the United States Department of Energy; and by the DSM CEA-Saclay.
  B.F. would like to thank the
CUFE for hospitality while the final part was done. His work is supported by Qiu-Shi
funding from Zhejiang University and Chinese NSF funding under
contract No.10875104.


\end{document}